\documentclass[structabstract]{aa}
\usepackage{graphicx}
\usepackage{epstopdf}
\usepackage{txfonts}
\usepackage{multirow}
\usepackage{longtable}
\usepackage[hmargin=2cm, vmargin=2cm]{geometry}

\begin{document}
   \title{High ions towards white dwarfs:  circumstellar line shifts and stellar temperature}


   \author{R. Lallement \inst{1}
   	\and
          B.Y. Welsh \inst{2}
           \and
           M.A. Barstow \inst{3}
           \and
           S.L. Casewell \inst{3}
         }

   \institute{  1 - GEPI, Observatoire de Paris, CNRS, 92195 Meudon, France\\
                \email{rosine.lallement@obspm.fr}\\
            2 - Space Sciences Laboratory, University of California, Berkeley, CA94720, USA \\
            3-  Department of Physics \& Astronomy, University of Leicester, University Road, Leicester,
LE1 7RH, UK}

   \date{Received ; revised }


\abstract
   {}
   { Our aim is to gain new insights into highly ionized gas towards nearby hot white dwarfs (WDs). The detection of absorption lines of highly ionized interstellar (IS) species in their spectra is the main diagnostic tool of the hot IS gas phase. This requires disentangling IS ions from photospheric or circumstellar (CS) ions, and thus their simultaneous study.}
   {We present far UV  spectra of three nearby white dwarfs recorded with the Cosmic Object Spectrograph on board the HST.} 
   {


Based on the compilation of OVI, CIV, SiIV and NV data from IUE, FUSE, GHRS, STIS, and COS, we derive an anti-correlation between the stellar temperature and the high ion velocity shift w.r.t. to the photosphere, with positive (resp. negative) velocity shifts for the cooler (resp. hotter) white dwarfs. This trend probably reflects more than a single process, however such a dependence on the WD's temperature again favors a CS origin for a very large fraction of those ion absorptions, previously observed with $\it IUE$, $\it HST$-STIS, $\it HST$-GHRS, $\it FUSE$, and now COS, selecting objects for which absorption line radial velocities, stellar effective temperature and photospheric velocity can be found in the literature. Interestingly, and gas in near-equilibrium in the star vicinity. It  is also probably significant that the temperature that corresponds to a null radial velocity, i.e. $\simeq$ 50,000K, also corresponds to the threshold below which there is a dichotomy between pure or heavy elements atmospheres as well as some temperature estimates for and a form of balance between radiation pressure and gravitation. This is consistent with ubiquitous evaporation of orbiting dusty material. Together with the fact that the fraction of stars with (red-or blue-) shifted lines and the fraction of stars known to possess heavy species in their atmosphere are of the same order, such a velocity-temperature relationship is consistent with quasi-continuous evaporation of orbiting CS dusty material, followed by accretion and settling down in the photosphere. In view of these results, ion measurements close to the photospheric or the IS velocity should be interpreted with caution, especially for stars at intermediate temperatures. While tracing CS gas, they may be erroneously attributed to photospheric material or to the ISM, explaining the difficulty of finding a coherent pattern of the high ions in the local IS 3D distribution.}

   {}

           
\maketitle
%

\section{Introduction}
Nearby hot white dwarfs (WDs) have long been used as background continuum sources of UV and EUV radiation for probing the absorption characteristics of the local (d $<$ 100pc) interstellar medium (Bruhweiler and Kondo 1983, Vennes et al 1996, Holberg et al 1999, Lehner et al 2003). Unfortunately the once assumed featureless UV continua of hot (DA) WDs have now been shown to be more complex than previously thought due to the presence of both photospheric and circumstellar lines (Bannister et al 2003). In particular the presence of relatively narrow stellar (and circumstellar) absorption  features due to the high ions of SiIV, CIV, NV and OVI has complicated the identification of similarly narrow interstellar features that are thought to be associated with hot (10$^{6}$K) gas assumed to be present within the confines of the Local Interstellar Cavity (Bannister et al 2003, Oegerle et al 2005, Savage and Lehner 2006, Barstow et al 2010). Observations also suggest that these high ionization species can be found in the absorption spectra of DA white dwarfs with surprisingly low (i.e. T $\sim$ 40,000K) effective temperatures (Holberg et al 1997, 1999). Thus, observing DA stars with relatively low (T $<$ 50,000K) effective temperatures may not necessarily guarantee that any detected high ionization features are not photospheric in their origin. 

Although radiative levitation and convection of heavy elements is a widely accepted mechanism for altering the stellar atmospheric contents of hot white dwarf  stars (Dreizler and Schuh 2001), other mechanisms have also been proposed to explain the presence of circumstellar absorption features. Bannister et al (2003) have proposed possible origins for these features that include ionization of the surrounding interstellar environment, mass loss in a stellar wind and the existence of material in an old planetary nebula around the stars. It has also been suggested that the accretion of gas from the surrounding interstellar medium may also play a significant role, at least for lower temperature white dwarfs (Koester and Wilken 2006). This has been refuted by Farihi et al (2010b), who propose that these white dwarfs are contaminated by circumstellar matter, the rocky remains of terrestrial planetary systems. Such accreting matter has recently been observed in the form of gas (and debris dust) disks around several nearby DZ white dwarfs (Kilic et al 2005, Farihi et al 2009, 2010a). This opens the exciting prospect of measuring the composition of circumstellar disks, and potentially of ancient extrasolar planetary systems (Zuckerman et al, 2007).

In principal, the 3 possible sources of absorption features in DA white dwarf UV spectra should be identifiable by an accurate measurement of their interstellar, photospheric and circumstellar velocities (V$_{ism}$, V$_{phot}$ and V$_{circ}$). This is especially true for nearby targets where the interstellar sight-lines have a simple absorption structure (i.e. few intervening interstellar clouds).  However, the vast majority of interstellar UV observations of nearby hot white dwarfs have been recorded with modest spectral resolution (R $\sim$ 20,000), such that it has proven difficult to distinguish between interstellar and non-interstellar features that lie within velocities $<$ 15 km s$^{-1}$ of each other. 

An accurate estimate of the interstellar velocity (V$_{ism}$) seen towards DA white dwarfs is normally gained through measurement of a (low ionization) UV resonance line such as OI $\lambda$1039\AA\, SII $\lambda$ 1260\AA\  or MgI $\lambda$ 2851\AA\ . However, it is often claimed that high ionization interstellar lines, formed at conductive interfaces, can be formed with a  significant velocity shift away from that of the low ionization species (Savage and Lehner 2006). Hence, although knowledge of V$_{ism}$ may be informative for the colder and low ionization local gas, in many cases this does not help in determining whether a high ionization feature is of local  interstellar origin. Theoretical models of the absorption due to the local complex of interstellar clouds ( d $<$ 5 pc) predict the presence of detectable levels of interstellar CIV and OVI (Ballet et al, 1986, Slavin 1989, Borkowski et al, 1990, Dalton and Balbus, 1993, Slavin and Frisch 2002, Gnat et al, 2010). However, although CIV absorption detected towards the B2Ve star HD 158427 has been associated with a most probable origin within the local cloud complex (Welsh et al 2010), we know of no similar claims for OVI or CIV features recorded towards local hot WDs. However, such detections of one or a few interfaces require excellent signal-to-noise ratios that are hard to achieve in the UV.

\begin{table*}[htdp]
\caption{Target stars, COS and FUSE observations}
\begin{center}
\begin{tabular}{|l|c|c|c|c|c|c|c|c|c|c|}
\hline
STAR & l$\_{II}$($^{\circ}$)& b$\_{II}$($^{\circ}$)& d(pc)  & Teff  &V$_{phot}$& COS G130M/G160M & FUSE/Prog\\
\hline
 WD2257-073 & 65.17 & -56.93 & 89$^{1}$ &  38000$^{3}$ & -7.8(+8.4,-7.5)* &2010/09/30 2344/1652s & AO5410100 2000/06/28  \\
  WD1942+499& 83.08&  12.75 & 89$^{2}$  & 33500$^{3}$ & 7.5($\pm$5)** &2009/10/28 502/828s &  Z90331010 2002/05/08 \\
WD1040+492 & 162.67& 57.01 & 230$^{1}$  & 47500$^{3}$  & -13.9($\pm$5)** &2010/01/06 720/1260s & Z004010010 2002/04/04 \\
 \hline

 \end{tabular}
\end{center}
\begin{footnotesize}

(1) Vennes et al., 1997
 (2) Vennes et al., 1998
(3) Barstow et al., 1994
(*) this study)
(**) from Barstow et al (2010)
\end{footnotesize}
\label{default}
\end{table*}%

\begin{figure*}[h]
	\begin{center}
\includegraphics[width=14cm]{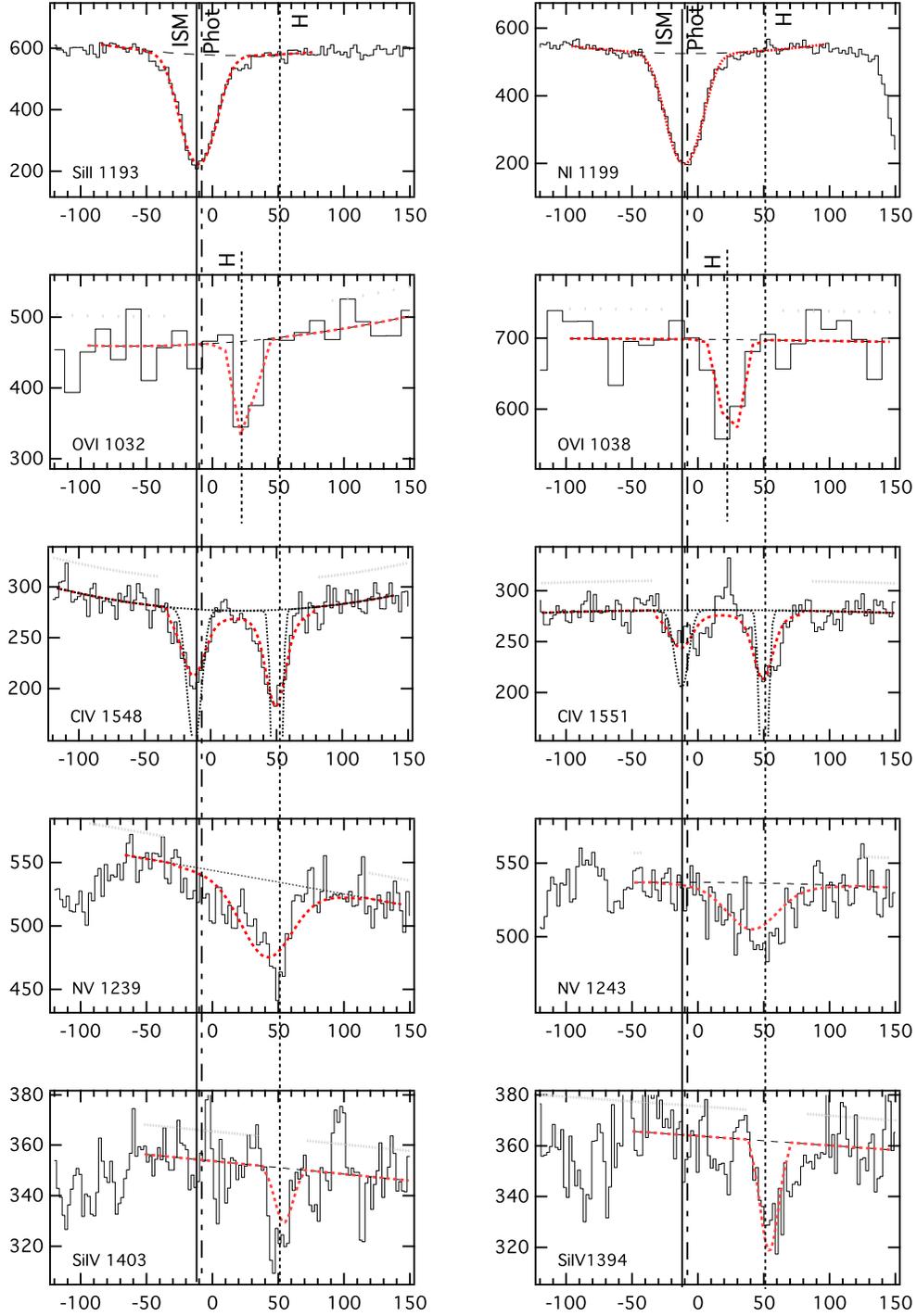}
\caption{Low and high ions in the COS spectra of WD2257-073. Also shown are the OVI lines as observed by FUSE. All velocities are heliocentric. Red doted lines are Voigt fits to the detected absorptions. The solid vertical line corresponds to the centroid of the cold-warm ISM absorption as traced SiII. The dotted line corresponds to the centre of the SiIV, NV and CIV lines. The dot-dot-dashed line corresponds to the Bar10 photospheric velocity.}
\end{center}
\end{figure*}

\begin{figure*}[h]
	\begin{center}
		\includegraphics[width=14cm]{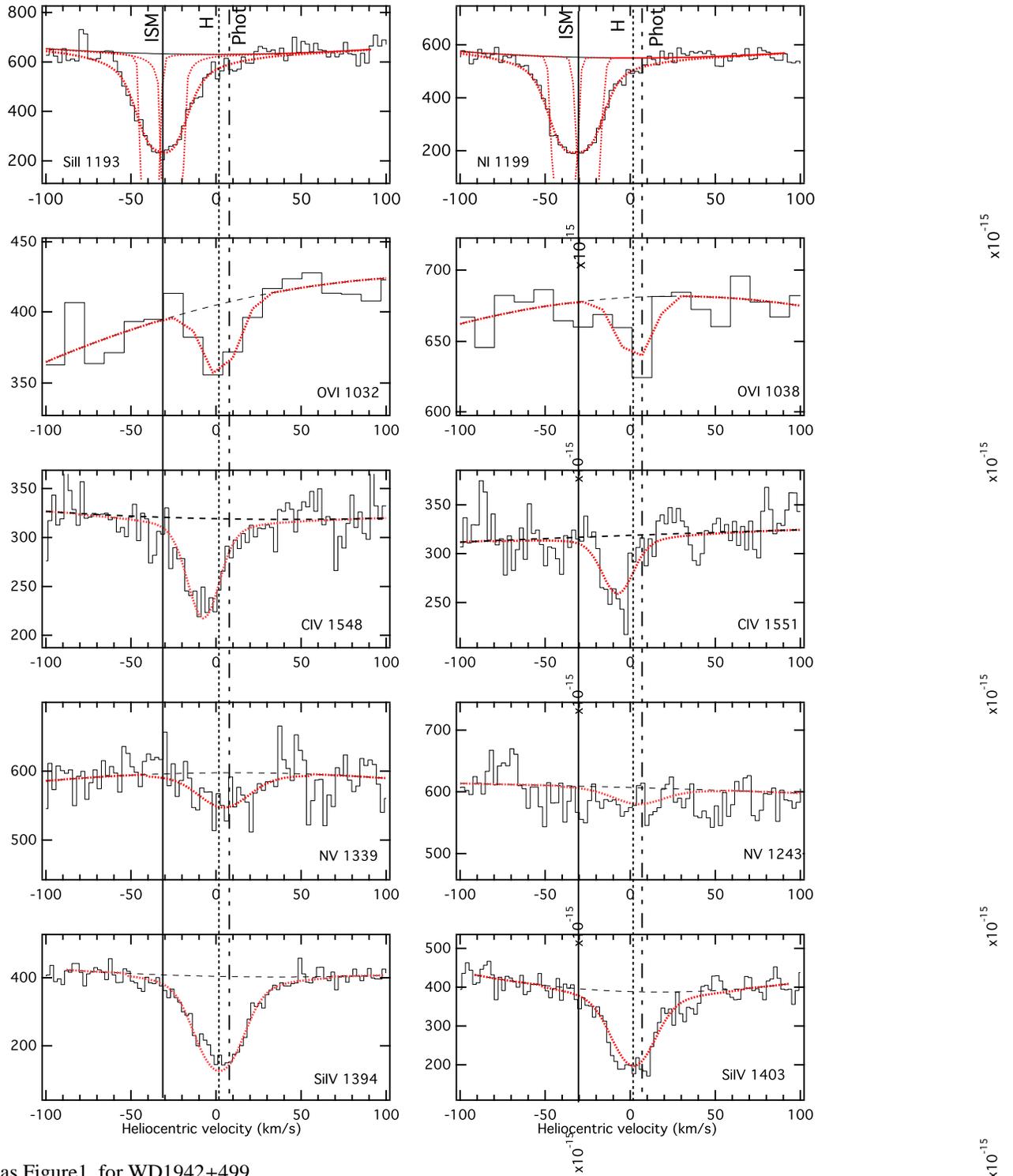}
\caption{Same as Figure1, for WD1942+499}
\end{center}
\end{figure*}

\begin{figure*}[h]
	\begin{center}
		\includegraphics[width=14cm]{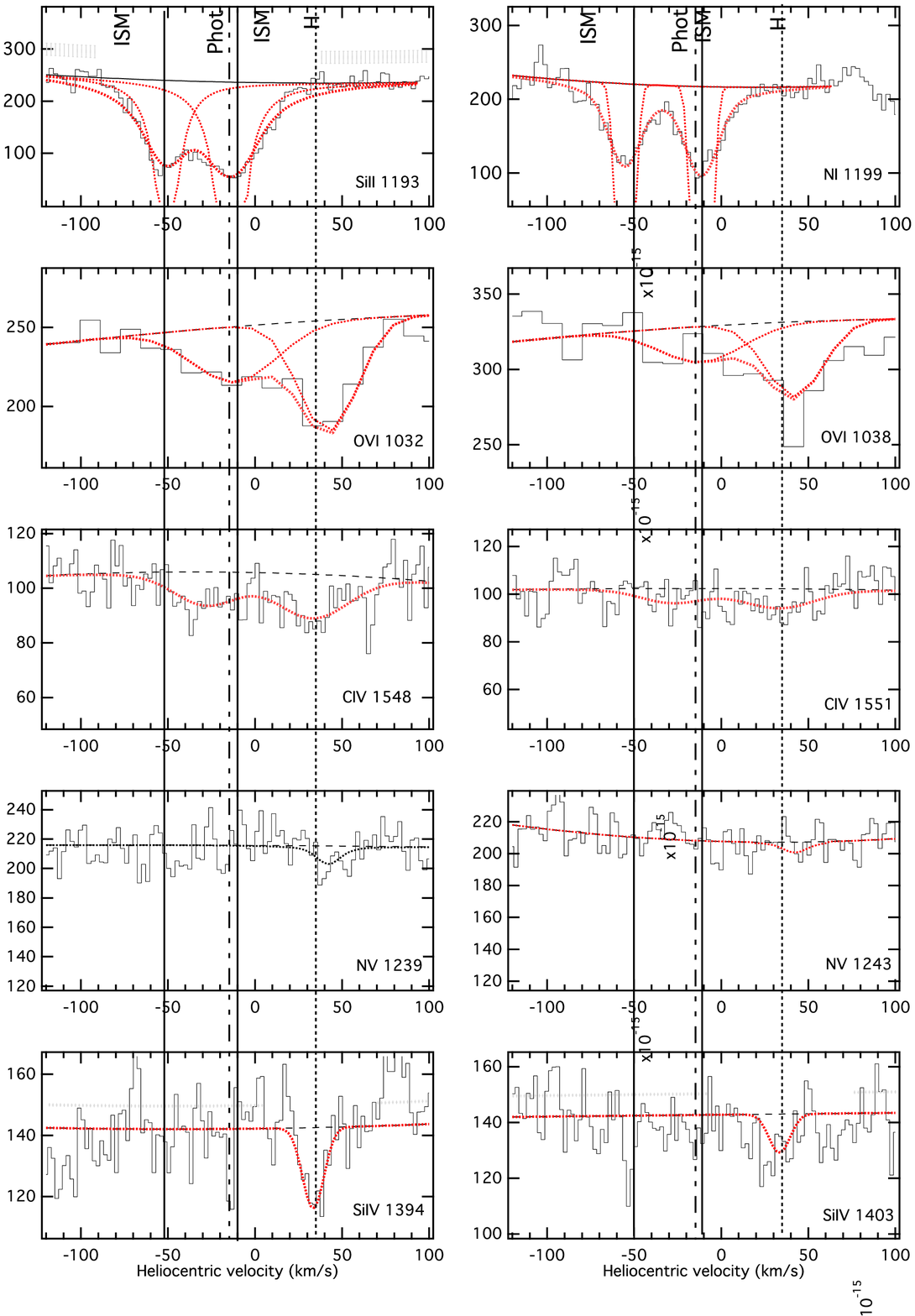}
\caption{Same as Figure 1, for WD1040+492.}
\end{center}
\end{figure*}



An accurate value for the V$_{phot}$ of a DA white dwarf can be obtained from measurement of the velocity of known photospheric lines that ideally possess wavelengths close to that of the high ionization line of interest (Savage and Lehner 2006). This technique generally works well for far UV spectra $<$ 1150\AA , but similarly useful photospheric indicators can be found in the  1200 - 1700\AA\ region (see Holberg, Barstow and Sion, 1998). 

Finally, values for V$_{circ}$ have normally been obtained under the assumption that the origin of such absorption features cannot be explained by photospheric or interstellar means. That is, their measured velocity is inconsistent with that obtained for both V$_{phot}$
 or V$_{ism}$ (Bannister et al 2003). This is a useful form of argument $\it only$ if one has great confidence in the values determined for V$_{phot}$ and V$_{ism}$, both of which (as argued previously) may  have significant uncertainties in their values.

An extensive compilation and reanalysis of Far Ultraviolet Spectroscopic Explorer (FUSE) OVI absorption data has been presented recently by Barstow et al (2010), hereafter Bar10, together with a new determination of the photospheric velocities, whenever possible, based on the good detections of heavy ion lines, if present, or on Lyman lines in their absence. 

As previously found (e.g. Savage and Lehner, 2006), for a fraction of the OVI detections the measured velocity is significantly shifted from the IS velocity (deduced from low ionization species), and another fraction of detections has no or
 negligible velocity shifts. In that work, the authors have also attempted to find some coherence between the OVI detection (or absence of detection) and the star location in the 3D distribution of the interstellar matter in the Sun vicinity. What was found is that the star location in the nearby cloud distribution seems to have no role, with detections being apparently randomly distributed on the sky, and thus reinforcing previous conclusions about the "patchy" nature of the OVI bearing gas.

In this Paper we present new UV absorption observations of three local DA white dwarfs using the Cosmic Origins Spectrograph on the HST (Osterman et al 2010). By combining these data with archival spectra gained with the FUSE satellite, and by carefully comparing the velocity shifts derived from a consistent set of low and high UV ion profiles recorded along each of the three sight-lines, we attempt to discriminate between the possible photospheric, circumstellar or interstellar origin of high ionization absorption features seen in these DA spectra. 

In a second step, we present an analysis of the high ion absorption towards all white dwarfs for which good quality data has been recorded, and for which estimates of both the stellar effective temperature and the photospheric radial velocity are available from the literature. In doing so, we make an extensive use of the recent compilation and spectral reanalysis presented by Bar10. Our goal is to search for a potential link between the
absorption line velocity shifts with the star characteristics that could shed light on the origin of the ions. 
The underlying idea is the following: if a large number of absorption lines are of interstellar origin, velocity shifts between the high and the low IS ions must be due to cloud/hot gas interface velocity gradients, as mentioned above, and no link should be found between the observed velocities and the stellar parameters. Conversely, if most lines have a circumstellar origin, one may expect a relationship between the star properties and the line shifts. 

The COS data are presented in section 2, and their analysis along with the FUSE data in section 3. In section 4 we show the results of the analysis of the archival data compilation. Section 5 presents some conclusions and a discussion on the interstellar, stellar and circumstellar contributions.


\section{Observations}

Observations of the 3 hot white dwarfs (WD2257-073, WD1942+499 and WD1040+492)  were made under the NASA HST Cycle 17 Cosmic Origins Spectrograph Guaranteed Time observation program GTO-11526, devoted to the study of nearby high ions of interstellar origin. The relevant astronomical information for the 3 targets have been taken from the Simbad data archive and are listed together with their exposure data in Table 1. The UV photon data were recorded through the 1 arc sec science aperture and, after reflection from either the G130M or G160M diffraction gratings, were incident on a micro-channel plate photon counting detector. The  COS data were extracted and processed using the latest version of the CALCOS pipeline software, with the final spectra presented here in the heliocentric velocity scale. The resolution of the resultant  spectra is $\sim$ 15 km s$^{-1}$. Typically, the S/N ratio for most of the COS data is $>$ 25:1.

We have also used FUSE far UV (912 - 1170\AA) spectra of all 3 targets, with the extracted datasets being downloaded from the Multi-Mission Archive at the Space Telescope Institute (MAST). These far UV spectra have a  spectral resolution comparable to that of the COS data, but are of lower S/N ratio of $\sim$ 15:1.

The three target stars were originally chosen for their very different locations w.r.t. the local interstellar cavity. WD2257-073 and WD1942+499 are very nearby (90 pc), but the former is at high latitude while 
the latter is close to the galactic plane. WD1040+492 is more distant (240 pc) and within the so-called {\it Local Chimney} that connects to the northern galactic halo.

\section{Data analysis}

Figure 1 shows the spectral regions centered on the high ions transitions $\lambda\lambda$1548.20/1550.77 CIV , $\lambda\lambda$1393.76/1402.77 SiIV , $\lambda\lambda$1238.82/1242.80NV. In order to allow an easy comparison of the high ions with the cold and warm ISM, we show at the same velocity scale  the $\lambda\lambda$1193.29 SiII and $\lambda\lambda$1199.55 NI transitions. The FUSE archival observations of the $\lambda\lambda$1036.613/1037.921 OVI doublet are also shown for comparison. 


COS and FUSE spectra have been fitted with model Voigt absorption profiles for the lines of SiIV , CIV , NV, SiII, NI and OVI. Figure 1 shows the fit results, which are detailed in Table 2.  For each transition we show the assumed continuum and the individual cloud absorptions. All doublets (SiIV, CIV, NV, OVI) have been fitted simultaneously to give a unique parameter set, while taking into account the noise level around both transitions. In some appropriate cases of strong lines and good signal we show in addition the absorption lines before convolution with the COS PSF. Details about the line-fit methodology can be found in Welsh and Lallement (2010).

Obtaining a meaningful comparison between the velocity shifts derived from the FUSE and COS  absorption lines is of great importance for the present study. 
This is problematic, because the accuracy of the FUSE wavelength scale is limited to $\simeq$ 15 km.s$^{-1}$. For the OVI and low ions line shifts obtained from
FUSE data,  Barstow et al (2010) used the $\lambda\lambda$ 1036.34 CII and $\lambda\lambda$1039.23 OI interstellar lines that are close to OVI doublet.
They then based their study on the relative difference between these low ions and OVI, a difference which is independent on the absolute wavelength scale. Here we propose to solve for the inter-calibration of the data, by using both FUSE and COS  low ionization IS lines as references, under the very reasonable assumption that the centroids of the strongest of those lines must be the same for both instruments. For the COS data we use the strong $\lambda\lambda$1193.29 SiII and $\lambda\lambda$1199.55 NI transitions. In practice we found that no calibration change was necessary, either by chance or because the calibration was better than
previously quoted. Table 3 shows the measured velocities for the low ionization species of CII and OI  based on FUSE measurements, and 
named V$_{ism}$ in Bar10,  as well as the COS velocities of the SiII and NI lines derived from our present profile-fitting. The differences
in velocities are smaller than 5 km.s$^{-1}$ for all lines, demonstrating that the wavelength scales agree rather well. Based on the comparison in Table3, we can reasonably assume that the difference between the FUSE velocity scale presented in Barstow et al (2010) and the COS scale is less then 5 kms$^{-1}$ for the three stars. Having kept the Bar10 velocity scales unchanged, our fit results for the OVI doublet are similar to their results, based on the same data. This allows for a
comparison of the OVI line shifts from FUSE data and the high ion line-shifts from COS data.  

 As detailed below, each of the three stars corresponds to a different situation as far as absorption from the high ion lines is concerned. For each star, we start by assuming
that a high ion line velocity close to the low ionisation IS velocity V$_{ism}$ is probably of interstellar origin. However, we also discuss the possibility of a significant velocity gradient among the IS lines due to conductive interfaces between the hot and warm phase of the ISM. 

\begin{table*}[htdp]
\caption{High ion fit results. Column-densities are in cm$^{-2}$ and velocities in km.s$^{-1}$}
\begin{center}
\begin{tabular}{|l|c|c|c|c|c|c|c|c|c|c|c|}
\hline
STAR & N(CIV)& V(CIV)& N(SiIV)& V(SiIV)  & N(NV)& V(NV) & N(OVI)& V(OVI) \\
  WD2257-073 & 1.0$\pm$0.4E13$^{*}$ & -12$\pm$1&  &   &   & &  &   \\
                           & 2.2$\pm$1.3 E13$^{**}$ & 50$\pm$1 & 1.1$\pm$0.5E12 & 54$\pm$2  & 1.2$\pm$0.3E13 & 41$\pm$2 & 3.6$\pm$1E13$^{1}$ & 25$\pm$1.5  \\
 WD1942+499& 1.6E13$^{7}$ &  -7$\pm$1 & 2.7$\pm$0.3E13$^{9}$ & 2$\pm$1 & 7.4$\pm$4.5E12$^{8}$ & 5$\pm$3 &  9.$\pm$3.E12$^{2}$&  3$\pm$2 \\
WD1040+492 & 7.5$\pm$4.0E12$^{5}$ & -27$\pm$5 &  &  &  & & 2.4$\pm$0.6E13$^{3}$  & -11.5$\pm$7 \\
                           & 1.1$\pm$0.5E13$^{6}$ & 34$\pm$4 & 1.6$\pm1.1$E12$^{4}$& 33.5$\pm$1.5  & 3$\pm$2E12 &  42$\pm$5& 3.5$\pm$1.5E13$^{3}$& 42$\pm$2.5\\

\hline
 \end{tabular}
\end{center}
\begin{footnotesize}
Constraints on the effective temperature from line fitting tests: (*) T=3.4$\pm$0.4E4 K; (**) T=9$\pm3.5$E3 K; (1) T$\leq$9000K. The upper limit on the column density corresponds to the assumption that T$\geq$3000K; 
(2) T$\leq$45000K. The upper limit on the column corresponds to the assumption that T$\geq$3000K; (3) T$\geq$500.000K;  (4) T$\geq$60,000K;(5) T$\geq$35,000K; (6) T$\geq$40,000K;(7) T=4.0$\pm2.0$E4 K; (8) T=2.4$\pm1.6$E5 K; (9) T=2.8$\pm0.3$E5 K

\end{footnotesize}
\label{default}
\end{table*}%

\begin{table*}[htdp]
\caption{Fit results for the low (resp. intermediate) ionization species SiII and NI. Also listed are the central velocities for the CII and OI absorptions measured with FUSE (Bar10). The similarity between the four measurements shows that the COS and FUSE wavelength scales  here agree very well. Column-densities are in cm$^{-2}$ and velocities in km.s$^{-1}$. (::) means a highly uncertain column.}
\begin{center}
\begin{tabular}{|l|c|c|c|c|c|c|c|c|c|}
\hline
STAR & N(SiII)& V(SiII)& N(NI)& V(NI) & V(CII)/V(OI)  (*) \\
  WD2257-073 & 1.5E13:: & -10.6$\pm$0.5 & 5.8E14:: &  -9.9$\pm$0.5  & -10.2  \\
  WD1942+499& 7 E14:: & -38.6 $\pm$1.5  & 5.E14:: &-38.6 $\pm$3 & -27.8**  \\
                          &1.5E14:: &-25.6 $\pm$2.0  &2E14:: & -24.3 $\pm$3.5 & -27.8**  \\
 WD1040+492 & 4.0E14::  & -50.7 $\pm$1.0  & 4.0E14::  & -55.1$\pm$1.0 &  -47.9 \\
                             & 5.0E15:: & -14.0 $\pm$1.0 & 2.5 E14::  & -11.1 $\pm$1.0 & -12.6 \\

\hline
 \end{tabular}
\end{center}
\begin{footnotesize}
(*) from Barstow et al., 2010
(**) total absorption centroid
\end{footnotesize}
\label{default}
\end{table*}%

\subsection{WD2257-073}

This star is particularly interesting because it is nearby (d = 89pc), and at low latitude. The IS absorption central velocity, $\simeq$ -10 kms$^{-1}$, is the expected value in case the intervening medium belongs to the group of local cloudlets (see e.g. Redfield and Linsky, 2008, hereafter RL08). More precisely, the line-of-sight passes close to the boundary of the LIC and of the RL08 Vel and Cet clouds. The LIC and Vel clouds velocities projected towards the star are both -2.0 kms$^{-1}$, while the Cet cloud vector projection is -13.6 kms$^{-1}$, close to the presently observed value. This suggests that beyond this local complex there is no other cloud in front of the star, and that high ions from hot-warm interfaces are expected only between the local complex and the hot gas of the local cavity. For this star,  the values of V$_{phot}$ , whose determination is described below, and V$_{ism}$ are very close ($\delta$V=2 kms$^{-1}$), which unfortunately complicates our analysis. 

As shown by Bar10, only one OVI line is detected (see Figure 1), that is redshifted and separated
 from both V$_{phot}$ $\&$ V$_{ism}$ by $\simeq$ +30 kms$^{-1}$, ruling out a stellar origin. 
Bar10 did not detect any photospheric material in the atmosphere of this star and, therefore, were not able to assign a photospheric velocity for the star. As a result, they interpreted the OVI detection, which is only slightly redshifted, as interstellar. In general it is possible to obtain a measurement of the photospheric velocity from the narrow core of the H$\alpha$ line. However, WD2257-073 resides in a binary system with HD217411, a G4V star, and its Balmer lines are obscured by the much brighter companion. In the far-UV the flux from the companion is negligible (see Barstow et al. 1994) and the Lyman series lines of the white dwarf can then be used to obtain a velocity measurement. Unfortunately, the line cores of the FUSE observation are filled in by geocoronal emission. Here we obtained the radial velocity by cross-correlating a series of synthetic Lyman line profiles to the available lines (Ly $\beta$ through $\epsilon$), yielding a value of -7.5 (+8.4, -7.5) km s$^{-1}$. While the measurement uncertainly is significant due to the broad line wings, it confirms clearly that the detected OVI is not photospheric.
If OVI is interstellar, then the large redshift w.r.t. the IS velocity implies a very strong velocity gradient originating in an evaporating hot-warm interface, as mentioned above, which is unlikely but not precluded. 
This hypothesis is however also contradicted by the small value of the OVI line width (or equivalently the low effective temperature, see Table 2), a value too small to correspond to the thermal broadening associated to the ion state, augmented by both the velocity gradient and the outward (i.e. towards the hot gas) displacement of the fractional abundance maximum predicted by the conductive interface models (e.g. Dalton and Balbus, 1993, Gnat et al, 2010). A final possibility is that the OVI absorption is a circumstellar line. 

The new COS spectra show two distinct and well resolved CIV doublet lines, one
 centered at V$_{phot}$ ($\simeq$ V$_{ism}$), the second being strongly redshifted (see Figure 1). This redshifted component is also clearly detected in the two SiIV and NV doublet transitions at exactly the same velocity as CIV, i.e. V$_{circ}$ = +50 kms$^{-1}$, while
  there is no detected NV nor SiIV counterpart to the CIV line found at V$_{phot}$. We now analyze the consequences of the COS observations in combination with that of FUSE.
  
The blueward component of the double CIV detections, which coincides with the photospheric velocity, has no SiIV, NV nor OVI counterpart, 
and thus it would be natural to conclude that this material is photospheric.  Indeed, it is possible to measure the abundance of CIV by matching the observed line profiles for the blue component of the doublet to a grid of synthetic spectra. We utilise the techniques described in Barstow et al (2003). Briefly, we computed a series of H-rich stellar model atmospheres using the TLUSTY non-LTE code. Heavier elements (C, N, O, Si, P and S) were included and a grid of models computed for varying abundances spanning the minimum and maximum observed line strengths. Synthetic spectra were calculated using the SYNSPEC program for comparison with the observational data using the XSPEC fitting package. XSPEC follows a $\chi^{2}$ minimisation algorithm to determine the best fit to the data, from which we derive the abundance of each element and the associated statistical measurement errors. The C IV resonance lines are best matched by an abundance of $\simeq$ 1.0 x 10$^{-8}$$\pm$10$^{-9}$ (Figure 4 top). However, such an abundance of photospheric material should give a very strong C III 1175/1176\AA\ multiplet at the temperature of this star. It does not, as illustrated in Figure 4 (bottom), which shows the region of the COS spectrum around the C III multiplet overlaid with the model prediction. The absence of photospheric CIII strongly disfavors the photospheric origin for the blueward CIV line.
 
 We now consider an interstellar origin for CIV. First, based on the simple line-of-sight structure discussed above, we can safely rule out an IS origin for both CIV lines simultaneously, which would imply two interfaces with extremely different velocity gradients (and for the redshifted line, a non realistic strong gradient). Secondly, IS conductive interfaces are very unlikely for both components due to the following reasons:
 
 -there would be corresponding OVI absorptions much stronger than the detected line. As a matter of fact, according to interface models (e.g. Slavin 1989, Gnat et al, 2010) the predicted model column density, N(OVI), is $\simeq$ 5 to 10 times higher than the CIV column, i.e. significantly higher than what is measured. This is first evidence for the independence of OVI and the two CIV lines. 

 -the line width of the two CIV doublets is again, as for OVI, much smaller than expected from interface models
 
 -for none of the CIV lines is there a possible dynamical association with the detected OVI that can be due to an interface:  i) if the blueward CIV absorption at the V$_{phot}$ $\&$ V$_{ism}$ velocity was generated in a conductive interface, then because V$_{CIV}$ = V$_{ism}$  this implies the absence of velocity gradient induced by the interface ionization structure. As a consequence one would also expect the corresponding OVI to lie at the same velocity, which is not observed. ii) in the case of the redshifted CIV, the difference between OVI and CIV is in the wrong sense, because in all cases OVI is more shifted than CIV w.r.t. to the cloud (i.e. the low ionization species velocity V$_{ism}$). Also, it is hard to explain why OVI would have a specific shift while the three other high ions CIV, NV and SiIV all absorb at exactly the same velocity. 
 
 -finally, the CIV columns are one order of magnitude higher than predicted by interface models, and here we expect only one or at most two interfaces only due to the simple LOS. 
 
In conclusion, we are left with a CS origin for the two CIV lines (and thus for the associated redshifted SiIV and NV), or alternatively with a CS origin for the CIV, SiIV and NV redshifted lines and a photo-ionized IS region for the blueward CIV. As a matter of fact photoionization models explain the existence of narrow lines of high ions, and can predict larger columns  than interfaces.  Similar CIV narrow lines have already been observed in the nearby and distant ISM (e.g. Fox et al, 2003, Welsh and Lallement, 2005, 2010, Knauth et al, 2003).

The OVI line is not associated with the other detected ions, and thus it could be generated within a hot-warm interface with corresponding CIV, SiIV and NV lines falling below our detection threshold, or alternately it could arise in CS gas. As mentioned above, the first hypothesis is unlikely due to the line width, and hence the CS origin is more likely. In that case, it is interesting that the velocity difference between the OVI absorption and the redshifted CIV, SiIV, NV lines is 25 km s$^{-1}$. Such a difference is larger than the wavelength calibration error quoted above, and we believe it is therefore significant. This implies that the 25 km s$^{-1}$ smaller OVI velocity, measured 10 years prior to the COS observations, may reflect a temporal variability of the circumstellar absorption. In the disk accretion scenario, such variability is not surprising.

 \begin{figure*}[h]
	\begin{center}
\includegraphics[height=12cm]{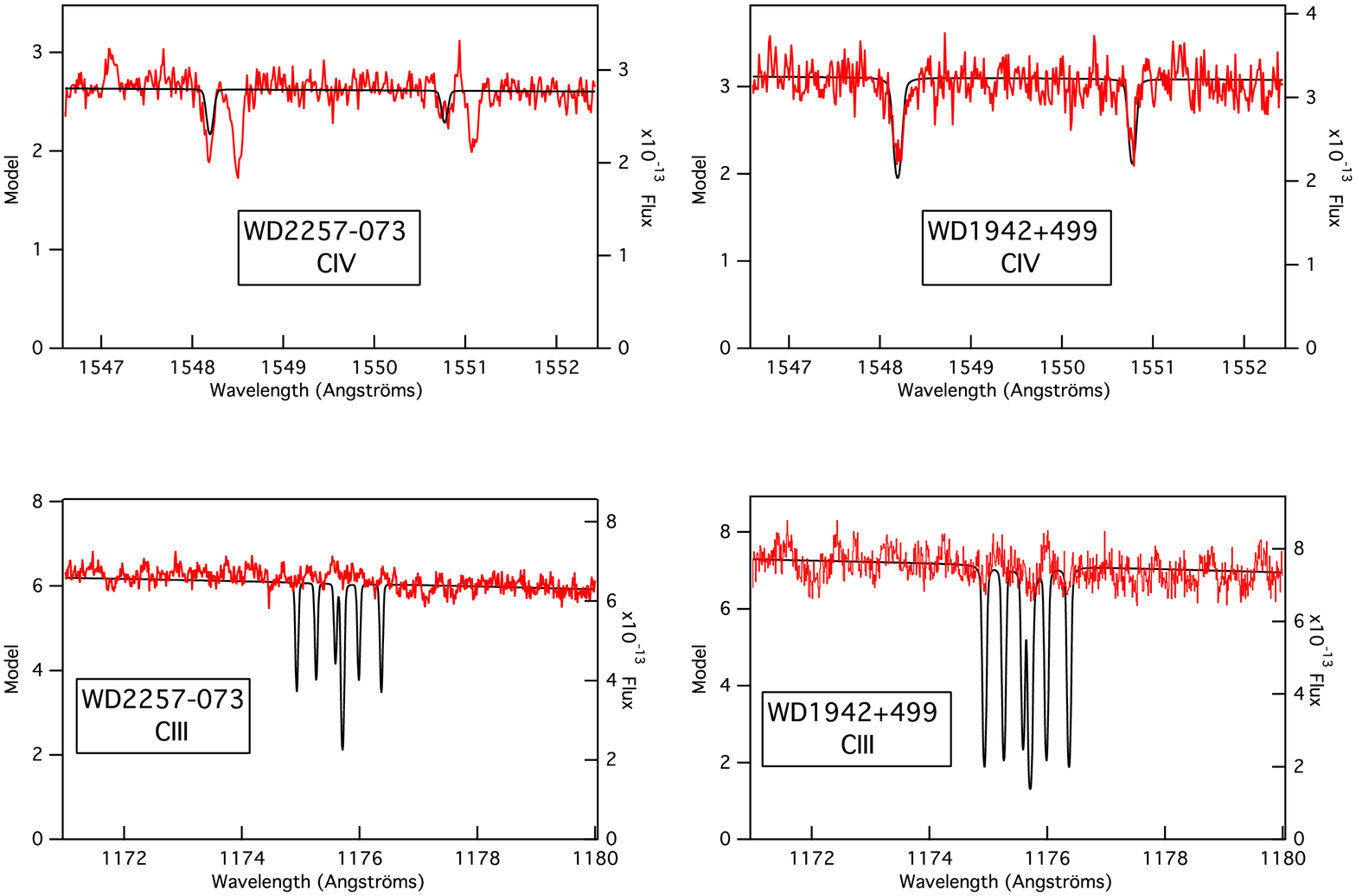}
\caption{Section of the COS spectrum of the white dwarfs WD2257-073 (left) and WD1942+499 (right) spanning the C IV doublet at 1548\AA\ and 1551 \AA\ (top) and the C III multiplet near 1175\AA\ (bottom). The data are overlaid with a synthetic spectrum calculated for an abundance that gives the best fit to the C IV doublet. For the two CIV lines that are observed at the photospheric velocity, i.e. the blueward line of WD2257-073 and the unique line of WD1942+499, photospheric models predict strong CIII multiplets that are not observed.}
\end{center}
\end{figure*}

 

 

\subsection{WD1942+499}

 WD1942-499 is an intermediate temperature hot DA white dwarf with Teff = 33 500K, also included in the FUSE study of O VI in the local ISM (Bar10). The far UV spectrum reveals detections of Si IV, P V and O VI, with the latter attributed to the photosphere of the star since the radial velocity of the line is close to that of the Si and P lines and differs significantly from the measured interstellar velocity (See Bar10). A search through the COS spectrum reveals detections of a number of low ionization species associated with the ISM (C II, O I, N I, S II, Si II, Fe II) and high ionization lines that are likely to be photospheric (C IV, Si III, Si IV). The radial velocities of the interstellar and photospheric lines are in mutual agreement, with mean values ~-27.8 km s$^{-1}$ and 7.5 km s$^{-1}$ respectively. 
 
 WD1942+499 is located close to the plane at a distance of 89pc, and is thus a good target to detect IS cloud interfaces associated with hot gas within the Local Cavity.  It is also of particular interest due to the large difference ($\simeq$ 40 km.s$^{-1}$) between its  photospheric velocity and the velocity of the strong interstellar absorbing cloud. According to RL08 this line-of-sight does not cross a specific cloud from the local cloud complex, but instead is close to the boundary of several clouds. The gas column intersected close to the Sun may be very small. The projected velocities of those nearby clouds span a range of  -8 to -19 km s$^{-1}$, which does not correspond to the center of the strong IS line revealed in our UV observations. This implies that the stronger absorption is generated farther away, probably close to the star itself, according to the maps of Vergely et al (2010) that do not reveal dense gas at closer distance. On the other hand, the local gas projected velocity is close to the phtospheric velocity.


For this star we do not detect any absorption close to V$_{ism}$ for the three high ions, in agreement with the lack of OVI at V$_{ism}$ (Bar10). On the other hand, CIV, SiIV and NV doublet lines are detected close to the photospheric velocity, and appear as the counterparts to the OVI doublet (Bar10). We also note that the ionized line of SiIII is also detected at V$_{phot}$, but not at V$_{ism}$. 

We consider first a photospheric origin for the high ions. We have analysed the detections of photospheric material in the manner described previously for WD2257-073. The C IV resonance lines are best matched by an abundance of $\sim$ 2.4 x10$^{-8}$$\pm$ 3 x 10$^{-9}$ (Figure 4, top right). However, as in the case of WD2257-073, such an abundance should give a very strong C III 1175/1176 \AA\ multiplet but none of these lines are present. Figure 4 (bottom right) shows the region of the COS spectrum around the C III multiplet overlaid with the model prediction. This result is difficult to understand. It is possible that this is a circumstellar component similar to those discovered in a number of hotter white dwarfs by Bannister et al (1995). A closer examination of Figure 2, shows that the centroid of both lines of the CIV doublet are shifted to the blue of the photospheric velocity, supporting this view. We note that Si III and Si IV abundances (~9.8 x 10$^{-7}$ $\pm$1.2 x 10$^{-7}$ and 2.2 x 10$^{-7}$ $\pm$ 1x 10$^{-7}$ respectively, measured from the FUSE spectrum) are discrepant indicating the presence of additional absorbing material not taken into account by a photospheric model. Also, we do not detect any oxygen in the COS spectrum. It is surprising that O V 1370\AA\ is not present, given the detection of O VI in the FUSE spectrum. This casts some doubt on the interpretation of the nature of the O VI line by Bar10, and our present  conclusion is that probably the high ion lines are not photospheric. 

For a number of reasons identical to our previous arguments for WD2257-073, an IS origin is also very unlikely. 
Firstly, all columns are far above what is expected for a hot-warm single interface, which must be the situation here because there is no velocity gradient (all ions are found at about the same velocity) and there can be only the interface with the local cloud complex. Secondly, the strength of OVI is far too small compared with model expectations and the measured CIV column. Thirdly, CIV and OVI line widths are much narrower than model predictions. One can also finally add the very high SiIV/CIV ratio that does not match the predictions of any of the models. 

In conclusion , we are left with a circumstellar source region for the detected high ions. Interestingly in this case, the CS lines are only very slightly shifted w.r.t. the photospheric velocity.




\subsection{WD1040+492}

 
 This target is the most distant and is located at high latitude in the so-called Local Chimney (Welsh et al, 1999) that connects the local cavity to the lower galactic halo. The line of sight intersects the local cloud complex and additional clouds at larger distances, as indicated by the observed complex velocity structure. Two main clouds are found at $\simeq$ -50 and -12 km s$^{-1}$. The absorption due to the local cloud(s) arises between 6 and 15 km s$^{-1}$ (LIC and NGP cloud, RL08).

 
 Two distinct OVI absorption components  have been measured by Bar10 towards WD1040+492, one
 centered at V$_{phot}$ (which is separated from V$_{ism}$ by only 15 km.s$^{-1}$), and a second component which is redshifted at +45km.s$^{-1}$. 
Our new COS data reveal, similarly to WD2257-073, CIV and SiIV counterparts to the redshifted OVI component , and a weak NV line. If  this redshifted component was interstellar, no velocity gradient is present, because all ions are aligned in velocity. Since there is no IS cloud at this velocity, an interface origin can be reasonably discarded. Indeed,  both the ratio between the OVI and CIV columns and the CIV column value also strongly disfavour an interstellar interface as the source region.  A circumstellar origin is thus the most likely explanation. 

A significant CIV line is also detected close to V$_{phot}$ and appears as the counterpart to the blueward OVI line. No SiIV nor NV counterpart is found concerning this blueward CIV, as in the case of WD2257-073. For this star however, contrary to WD2257-073, the CIV/OVI ratio is marginally compatible with IS interface models, as is the upper limit on SiIV and NV.  The OVI and CIV velocity centroids do not perfectly match the strongest IS absorption component centers. However, the velocity structure is likely to be more complex than the idealized two-component model structure presented here, and there may be multiple interfaces. This multiplicity can also possibly  explain the high CIV column due to the presence of multiple interfaces.  We conclude here that there is possibly a detection of hot gas from IS interfaces, however, we can not dismiss a photospheric origin, due to the proximity between the photospheric and IS lines.


\section{High Ion radial velocities and stellar temperature}

\begin{figure*}[h]
	\begin{center}
\includegraphics[width=15cm,height=11cm]{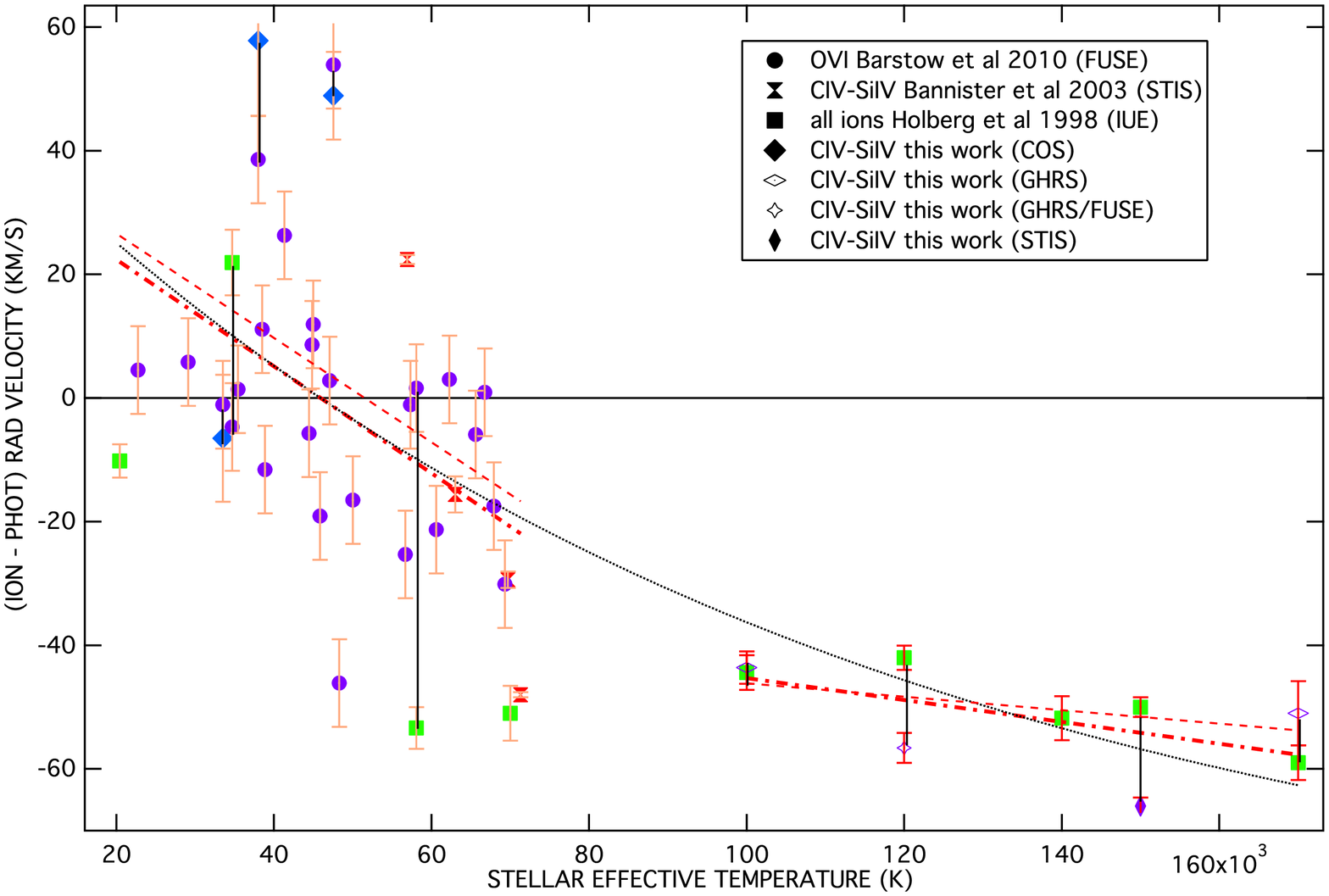}
\caption{High ion velocity shifts with respect to V$_{phot}$ as a function of the WD temperature: white dwarf FUSE, IUE and HST-STIS  data from Holberg et al 1998, Bannister et al 2003 and Bar10  as well as the present COS data have been combined and for all high ion detections their velocity centroid is shown as a function of the stellar effective temperature. In addition, HST and FUSE archive data have been analyzed for four of the five hottest stars, in order to provide a comparison with the IUE data. Temperatures are from Barstow et al (1994, 2001, 2003, 2010), Christian et al 1996, Finley et al 1997, Holberg et al 1999, Landsman et al 1993, Marsh et al 1997, Saffer et al 1998, Vennes et al (1997,1998), Werner et al, 1997. Photospheric velocities are taken from the sources indicated in the legend, with the latest Bar10 results being chosen in case of multiple determinations. In case of measurements with two different instruments towards the same target, a vertical line links the two data points. Despite the large scatter, there is a clear decrease of the relative velocity with the stellar temperature. An arbitrary exponential fit to all data is shown (black dotted lines). Also shown are linear fits for T $\leq$ 75,000K and T  $\geq$ 75,000K resp. Dotted lines are fits corresponding to the shown error bars. Dash-dotted lines are fits that use the overall dispersion as the mean error (see text). The temperature for which the average shift w.r.t the photospheric lines (resp. the star) is null is $\simeq$ 50,000 K.}
\end{center}
\end{figure*}



\begin{table*}[htdp]
\caption{High ions absorption line shifts and stellar temperatures. Error bars are taken from the referenced articles or estimated from the present COS measurements. They reflect uncertainties linked to continuum and line fitting and the quality of the signal. They do not include calibration uncertainties, that disappear in the calculation of the velocity shift.}
\begin{center}
\begin{tabular}{|c|c|c|c|c|c|}
\hline
STAR  & Teff(*)  &V$_{ion}$ & V$_{phot}$ & Shift  & Ref. /Inst.(**)\\
\hline
WD1337+701&20435&-10.5 $\pm$ 2.3 & -0.3 $\pm$ 1.4 & -10.2 $\pm$ 2.7 &1 I\\
WD0512+326&22750&25 $\pm$5 &20.5 $\pm$5 &4.5 $\pm$ 7 & 2 F\\
WD0252-055&29120 &16.2 $\pm$5 &10.4 $\pm$5 &5.8 V 7 & 2 F\\
WD1942+499&33500&6.4 $\pm$5 &7.5 $\pm$5 &-1.1 $\pm$ 7 & 2 F\\
" & " &1$\pm$ 7 &7.5 $\pm$ 5 &-6.5 $\pm$ 10 &3 C\\
WD0050-332&34684&13.7 $\pm$5 &18.4 $\pm$5 &-4.7 $\pm$ 7 & 2 F\\
" & " &40.3 $\pm$ 1.8 &18.4  $\pm$5 &21.9 $\pm$ 5 &1 I\\
WD1021+266&35432&13.4 $\pm$5 &12 $\pm$5 &1.4 $\pm$ 7 &2 F\\
WD2257-073&38010&50$\pm$ 7 &-7.8 (+13.6,-7.5)&57.8 $\pm$ 12 &3 C\\
" & " &30.8 $\pm$5 &-7.8 $\pm$5 &38.6 $\pm$ 7 &2 F\\
WD1611-084&38500&-24.2 $\pm$5 &-35.3 $\pm$5 &11.1 $\pm$ 7 &2 F\\
WD2111+498&38866&13.1 $\pm$5 &24.7 $\pm$5 &-11.6 $\pm$ 7 &2 F\\
WD1950-432&41339&0.3$\pm$5&-26$\pm$5&26.3 $\pm$ 7 &2 F\\
WD2000-561&44456&-15.3$\pm$5&-9.6$\pm$5&-5.7 $\pm$ 7 &2 F\\
WD0131-164&44850&19.5$\pm$5&10.9$\pm$5&8.6 $\pm$ 7 &2 F\\
WD1029+537&44980&25.7$\pm$5&13.8$\pm$5&11.9 $\pm$ 7 &2 F\\
WD2321-549&45860&-1.6$\pm$5&17.5$\pm$5&-19.1 $\pm$ 7 &2 F\\
WD2011+398&47057&13.9$\pm$5&11.1$\pm$5&2.8 $\pm$ 7 &2 F\\
WD1040+492&47560&35$\pm$ 5 &-13.9 $\pm$5&48.9 $\pm$ 7 &3 C\\
" & " &45$\pm$5&-13.9$\pm$5&53.9 $\pm$ 7 &2 F\\
WD2124-224&48297&-16.4$\pm$5&29.7$\pm$5&-46.1 $\pm$ 7 &2 F\\
WD0226-615&50000&9.3$\pm$5&25.8$\pm$5&-16.5 $\pm$ 7 &2 F\\
WD2331-475&56682&3$\pm$5&28.3$\pm$5&-25.3 $\pm$ 7 &2 F\\
WD2218+706&56900&-16.3 $\pm$ 0.7 &-38.7 $\pm$ 0.2 &22.4 $\pm$ 0.7 &4 S\\
WD0501+524&57340&20.4$\pm$5&21.5$\pm$5&-1.1 $\pm$ 7 &2 F\\
WD0455-282&58080&16.2 $\pm$ 2.7 &69.6 $\pm$ 2.0 &-53.4$\pm$ 3.5 &1 I\\
" & " &59.3$\pm$5&57.7$\pm$5&1.6 $\pm$ 7 &2 F\\
WD0027-636&60595&0.5$\pm$5&21.8$\pm$5&-21.3 $\pm$ 7 &2 F\\
WD0621-376&62280&34.3$\pm$5&31.3$\pm$5&3 $\pm$ 7 &2 F\\
REJ0558-373&63000&7.1 $\pm$ 0.8 &22.7 $\pm$ 2.8 &-15.6 $\pm$ 3 &4 S\\
WD2211-495&65600&18$\pm$5&23.9$\pm$5&-5.9 $\pm$ 7 &2 F\\
WD1342+442&66750&-12.2$\pm$5&-13.1$\pm$5&0.9 $\pm$ 7 &2 F\\
WD2146-433&67912&10.4$\pm$5&27.9$\pm$5&-17.5 $\pm$ 7 &2 F\\
WD2350-706&69300&10.6$\pm$5&40.7$\pm$5&-30.1 $\pm$ 7 &2 F\\
Ton021&69700&7.6 $\pm$ 1.3 &37.1 $\pm$ 0.2 &-29.4 $\pm$ 1.3 &4 S\\
WD0501-289&70000&-12.2 $\pm$ 4.3 &39.3 $\pm$ 1.1 &-51.5 $\pm$ 4.4  &1 I\\
REJ1738+665&71300&-17.8 $\pm$ 0.3 &30.2 $\pm$ 0.3 &-48. 0 $\pm$ 0.4 &4 S\\
WD1034+001&100000&6.6 $\pm$ 2.5 &50.8 $\pm$ 1.3 &-44.4 $\pm$ 2.8 &1 I\\
 " &" & 7.9 $\pm$ 1.8 &  51.5 $\pm$ 1.9 & -43.6 $\pm$ $\pm$ & 3 G \\
WD0005+511&120000&-6.2 $\pm$ 1.0 &36.1 $\pm$ 1.7 &-42.3 $\pm$ 2 &1 I\\
 " &  " & -22.5 $\pm$ 1.4 F & 34.1 $\pm$2.0 G&  -56.6 $\pm$ 2.4 F/G & 3 F/G\\
WD1159-034&140000&-1.8 $\pm$ 3.4 & 50.1 $\pm$1.1 &-51.8 $\pm$ 3.6 &1 I\\
" &  " & und  & 47.4 $\pm$ 1 & und & 3 F\\
WD0044-121&150000&-70.2 $\pm$ 1.05 &-22.0 $\pm$ 1.2 &-50.2 $\pm$ 1.6 &1 I\\
" &  " &  -80 $\pm$ 1.1 &  -14 $\pm$ 0.8 & -66.0 $\pm$ 1.4 & 3 S\\
WD2117+342&170000&-11.9 $\pm$ 1.1 & 47.2 $\pm$ 2.6 &-59.1$\pm$ 2.8 &1 I\\
" &  " & -29 $\pm$ 2.4 & 22 $\pm$ 4.6 & -51 $\pm$ 5.2&  3 G\\
\hline
 \hline
 \end{tabular}
\end{center}
\begin{footnotesize}
(*) References for temperatures are listed in Fig. 5 caption.\\
(**) 1: Holberg et al, 1998, 2: Barstow et al, 2010, 3: this work, 4: Bannister et al, 2003\\
I=IUE;F=FUSE;C=COS;S=STIS;G=GHRS
\end{footnotesize}
\label{default}
\end{table*}%


Our analysis of the COS data has not allowed an unambiguous detection of local IS hot gas, but the combination of the FUSE and COS diagnostics has brought other results with new and important implications. Together they reinforce the frequency of high ion absorptions that are not of interstellar nor stellar origin, and are thus presumably circumstellar. Our data also show that the CS OVI lines are accompanied by CIV, NV or SiIV absorptions at similar velocities. The case of WD1942 also demonstrates that
CS absorptions can be found at velocities that are very close to the photospheric velocity. This has led us to investigate further the distribution of the high ion velocity shifts, and in particular to search for  possible statistical trends using a large set of targets. This may help overcoming the difficulties linked to the limited resolution, the uncertainties, the blending of lines and the oversimplification of the velocity structure 
when analyzing individual LOS data.  

We report here on a search for a statistical relationship between the stellar temperature and the velocity characteristics of the high ion absorption lines. To do so, we have greatly benefited from the large increase in the number of targets recently studied by Barstow et al (2010), who have presented a compilation and reanalysis of 80 lines-of-sight towards nearby DA WD stars. We have combined this data set with the SiIV, NV, CIV measurements reported by Bannister et al (2003) and Holberg et al (1998), mostly for hot WDs.  In addition, we have reanalyzed HST/GHRS and STIS archive data as well as FUSE data from the Holberg et al (1998) study of the hottest white dwarfs. From all sources we have retained those targets for which effective temperature and photospheric velocity have been determined (or estimated) and can be found in the literature. We have excluded the star WD0232+035 (Feige 24) since it is part of a binary system that is close enough to have undergone a common envelope phase. Thus, any observed circumstellar material may have a quite different origin to that detected towards an isolated white dwarf star. The other binaries (WD0512+326, WD0252-055, WD0226-615,WD2350-706
are all widely separated and can be treated as isolated objects in the context of this work. The total number of WDs for which these conditions are fulfilled amounts to  37 and the total number of measured velocity shifts amounts to 46. For the case of well separated OVI absorption line components we have used the  absorption with the largest velocity shift w.r.t. the photospheric velocity.   Table 4 lists all of the selected measurements, and Figure 5 shows for this dataset the velocity difference between the high ion velocity centroid and V$_{phot}$ as a function of the stellar temperature. For the cases ( 9 targets)
for which the OVI data and the other ions give separate results, we have treated those results separately and linked the corresponding data points in the figure.  Error bars for each source are taken from the references listed in Table 4. All Bar10 measurements are given a conservative error bar of 5 km.s$^{-1}$. The velocity shifts to the COS data are an average of the CIV, SIIV and NV values from Table 2 and are assigned an uncertainty estimated from the error bars and the velocity scatter among the ions. However, we note that the (small) differences in velocity for the various ions may be real and could be linked to their physical production mechanism. For all error bars we do  not take into account the wavelength calibration uncertainty, since ultimately this uncertainty disappears when calculating the V$_{ion}$-V$_{phot}$ velocity difference.

Figure 5 shows that in many cases there are strong differences between the shifts obtained from different measurements towards the same target, especially for different ions (e.g. OVI and CIV/SiIV), as we have found in our present analysis of the COS data. Smaller differences, still above error bars, are also found for the same ion (three cases among the very hot WD's). However, despite the strong scatter, Figure 5 shows that the relative velocity decreases with increasing stellar temperature, over the entire range from 20,000 to 170,000K. Hot stars display mostly negative values (outwards radial velocities) while the cooler stars reveal positive velocities, i.e. towards the star. 
A Spearman test suggests an anti-correlation at better than 99.98 \% when using all data points (46 shifts). 

The ensemble of targets we have used includes very different objects, and in particular the hottest white dwarfs studied with $\it I.U.E.$. However, we note that  the anti-correlation between the temperature and the average high ion radial velocity also holds when excluding the strongly negative values corresponding to these hottest dwarfs (from Holberg et al, 1998), namely temperatures above 70,000K. In this case the absence of an anti-correlation is rejected with 98.5 \% confidence for 37 shifts. In order to explore any possible relationship for the data in Figure 5 we start by 
showing simple linear fits for sources with T $\leq$ 75,000K and for those with T  $\geq$ 100,000K. The former slope is -0.85  $\pm$ 0.28 10$^{-3}$ km.s$^{-1}$K$^{-1}$, i.e. 3 times the standard deviation, while the high temperature slope obtained from 5 stars and 9 shifts, -0.18  $\pm$ 0.09 10$^{-3}$ km.s$^{-1}$K$^{-1}$ is only marginally significant. For those linear fits the error bars associated with the best-fit coefficients are deduced from the overall dispersion of the data points, i.e. the mean error is deduced from this dispersion. Linear fits making use of the error bars given in Table 5 are slightly different and shown as dotted lines. Because we believe that a large fraction of the dispersion in the fits is not linked to the data but does in fact reflect a real variability, the former fits are very likely more appropriate. The latter fits lead to an unrealistically small error bar on the fit coefficients. The difference between the two curves may reflect different mechanisms and physical states of circumstellar matter for the coolest and hottest objects. However, we note that there is a good continuity between the two temperature regimes of the data. When we extrapolate the low temperature line-fit towards the high temperature data regime, it intersects the high temperature fit close to the 100,000 K data points. We have also added an exponential fit to the whole data set, that has no physical basis, but aids the eye when comparing the two other fit relationships. In order to establish the actual dependence of the velocity shift on stellar temperature we shall ultimately require many more data points, especially at the low and high temperature extremes.

For most of the velocity shifts their large values preclude an interstellar origin, and by definition also a stellar origin, i.e. the discovered relationship therefore applies mostly to circumstellar absorptions. An influence of the stellar temperature on the gas motions very likely reflects the role of stellar radiation pressure. Such motions also imply gas falling onto or moving away from the WD star, i.e. a non stationary behavior which can be linked to gas evaporating from circumstellar dust or solid bodies. It is beyond the scope of this work to model the evaporation process, but we believe the global trend shown in Figure 5 may help in formulating such a model.  On the other hand (as discussed below), Figure 5 most likely contains both interstellar and photospheric lines in addition to CS absorptions, which produces an additional dispersion to the fitted data points.

\section{Discussion}

 While our COS observations were originally motivated by a search for IS absorptions, the results have guided us in a different direction, and have required a discussion in the broader perspective of stellar, circumstellar and interstellar ion production.

Overall  the three COS spectra, when analyzed in parallel with FUSE data, have not provided us
with definitive results on the location of hot gas in the vicinity
of the Sun. For only one star is there a potential detection of high ions originating in hot-warm interfaces. WD1040+492 is located at 240 pc and high above the Plane, in the soft X-ray emitting northern {\it chimney} that connects the LB to the halo (see Koutroumpa et al, 2009). This would confirm that hot, one million K gas is present closer than $\simeq$ 200 pc. On the other hand the question of the nature of the tenuous gas filling the LB close to the Plane remains unsolved, our detection threshold having not permitted to firmly detect hot-warm interfaces towards the two closest targets. We note that recent work by Peek et al (2011) has ruled out any soft X-ray emission from a local hot bubble to a very low level.
 
On the other hand, we believe we have shed new light on the general question of the origin of the high ion lines, potentially explaining why
their true nature has remained elusive despite the accumulation of much data over the years.
Firstly, and very surprisingly, all three targets appear to have circumstellar high ions. One of the targets may even display two CS lines. This strongly reinforces the ubiquity of such absorptions in the UV spectra of hot WDs.

The second result is associated with our compilation of high ion data in which we have found an anti-correlation between 
the high ion velocity shift and the stellar temperature. Such a trend is pointing towards their origin in the environment of the star, i.e. a stellar/interstellar interface or circumstellar matter.  Because  the shifts reach large negative or positive amplitudes, the hypothesis of an interstellar/stellar interface as the source of the ions is unlikely. Str\"omgren spheres exist around hot white dwarfs (Dupree and Raymond, 1983), but the ion velocities within those spheres should have intermediate values  between the IS velocity and the stellar velocity. On the other hand, the trend shown in Figure 5 and the
associated large velocity values provide strong evidence for a circumstellar origin for most of the observed high ion lines. In a very simple interpretation, the higher the star effective temperature, the stronger the stellar radiation pressure, and subsequently the more repulsive the force resulting from gravitation and radiation pressure. Material evaporating from a solid body circling around the star is rapidly ionized and subjected to a force that at first order is either attractive or repulsive depending on the balance between those two actions. Interestingly the temperature that corresponds to the smallest 
velocity shifts w.r.t. V$_{phot}$ occur for temperatures of the order of 40,000K-50,000K , i.e. close to the range of temperature for which metals can levitate into the stellar atmosphere. It is also the threshold below which a dichotomy is found between pure H atmospheres and "polluted" WD atmospheres.

This interpretation is certainly oversimplified and more data and extensive modeling are required to go beyond. Also, some of the shifts displayed in Figure 5 certainly do not correspond to CS material, but correspond to IS or photospheric lines, which introduces biases without a physical meaning. On the observational side, it may be objected that the ensemble of targets we have used includes very different objects, and in particular the hottest white dwarfs studied with $\it I.U.E.$. For those objects possessing strongly blue-shifted lines the physical processes at work may be totally different from those at play in the cooler objects. In particular they comprise hot DO or PG1159 stars that may have had recent mass-loss and planetary nebulae associated with them. Moreover, blue-shifted lines from most of the Bannister al al sample may be related to outflow of material close to the WD surface (i.e. deep in the gravitational potential well of the WD) or material located at some distance from the surface (i.e. not as red-shifted by the gravitational field), or some combination of these. This is why we have treated the T$\leq$ 70,000K and T$\geq$ 70,000K objects separately. The resulting linear fits in Fig 5 have different slopes and may indeed reflect very different underlying physical mechanisms. We note that the anti-correlation holds when excluding the very hot white dwarfs analyzed by Holberg et al (1998), suggesting that for those objects the majority of the absorbing material is strongly influenced by the stellar environment in a kind of global manner leading to the observed trend, even if more than one process is present.  On the other hand, for the very hot white dwarfs the data are marginally compatible with a temperature-independent velocity shift. In order to test the actual physical processes at play, realistic models of the physical and dynamical properties of evaporating material are required. This is far beyond the scope of the present work. Here we are simply emphasizing the need for such appropriate models of evaporation-accretion/ejection of CS matter that can potentially explain the newly found V-Teff relationship.

The third result, from Figure 5, is that the absorbing material at the origin of the V-Teff trend, which is linked to the star itself, may correspond to moderate and low velocities  (say, between -20 and +20 km.s$^{-1}$). As a consequence a large fraction of the data in Figure 5 is likely to be associated with the same CS phenomenon, despite the small velocity shifts. Due to the similarity in observed velocity of both photospheric and interstellar lines, their true nature may have previously been incorrectly attributed to IS or photospheric material. This would lead to an erroneous interpretation of the data and may explain the present unclear view of the spatial distribution of any hot gas (see e.g. Bar10, Savage and Lehner, 2006). Disentangling those absorptions from the true IS lines is difficult but more than ever necessary, and requires high resolution and high quality absorption data. WD targets were initially chosen for their supposedly featureless continua, but due to the accretion processes they may not be the most convenient targets. There is only a small amount of similar
data taken towards B stars. Welsh and Lallement (2008) failed to detect OVI towards any star closer than
80pc, lowering the upper limit on the LB hot gas pressure. More recently, CIV has been detected within the Local cavity and closer than 74pc using STIS observations of B stars (Welsh et al, 2010). The CIV narrow Doppler width suggested that  these ions do not belong to a cloud/hot gas interface, but rather formed  by photoionization. However, nearby B-stars located within the local ISM are often
too bright to be observed with spectrographs on HST.

Finally, we would like to emphasize that the recent observations of dusty disks and the prospect of measuring abundances of ancient extrasolar systems around white dwarfs requires a better understanding of all the chain of processes, including disk material evaporation. More numerous and precise measurements of the high ion absorptions may also help understanding this particular step of the whole process.


\begin{acknowledgements}
We thank our anonymous referee for useful suggestions. MAB was supported by the Science and Technology Facilities Council, UK. SLC was supported by a University of Leicester post-doctoral research position.

\end{acknowledgements}


\begin{thebibliography}{}


\bibitem[Ballet et 
al.(1986)]{1986A&A...161...12B} Ballet, J., Arnaud, M., \& Rothenflug, R.\ 1986, \aap, 161, 12 

\bibitem[Bannister et al.(2003)]{2003MNRAS.341..477B} Bannister, N.~P., 
Barstow, M.~A., Holberg, J.~B., 
\& Bruhweiler, F.~C.\ 2003, \mnras, 341, 477 

\bibitem[Barstow et al.(1994)]{1994MNRAS.270..499B} Barstow, M.~A., 
Holberg, J.~B., Fleming, T.~A., Marsh, M.~C., Koester, D., 
\& Wonnacott, D.\ 1994, \mnras, 270, 499 


\bibitem[Barstow et al.(2001)]{2001MNRAS.328..211B} Barstow, M.~A., 
Holberg, J.~B., Hubeny, I., Good, S.~A., Levan, A.~J., 
\& Meru, F.\ 2001, \mnras, 328, 211 

\bibitem[Barstow et al.(2003)]{2003MNRAS.341..870B} Barstow, M.~A., Good, 
S.~A., Holberg, J.~B., Hubeny, I., Bannister, N.~P., Bruhweiler, F.~C., 
Burleigh, M.~R., \& Napiwotzki, R.\ 2003, \mnras, 341, 870 

\bibitem[Barstow et al.(2010)]{2010ApJ...723.1762B} Barstow, M.~A., Boyce, 
D.~D., Welsh, B.~Y., Lallement, R., Barstow, J.~K., Forbes, A.~E., 
\& Preval, S.\ 2010, \apj, 723, 1762  (Bar10)

\bibitem[Borkowski et al.(1990)]{1990ApJ...355..501B} Borkowski, K.~J., 
Balbus, S.~A., \& Fristrom, C.~C.\ 1990, \apj, 355, 501 

\bibitem[Bruhweiler 
\& Kondo(1983)]{1983ApJ...269..657B} Bruhweiler, F.~C., \& Kondo, Y.\ 1983, \apj, 269, 657 


\bibitem[Christian et al.(1996)]{1996AJ....112..258C} Christian, D.~J., 
Vennes, S., Thorstensen, J.~R., \& Mathioudakis, M.\ 1996, \aj, 112, 258 

\bibitem[Cox(1998)]{1998LNP...506..121C} Cox, D.~P.\ 1998, IAU Colloq.~166:
The Local Bubble and Beyond, 506, 121

\bibitem[Dalton 
\& Balbus(1993)]{1993ApJ...404..625D} Dalton, W.~W., \& Balbus, S.~A.\ 1993, \apj, 404, 625 

\bibitem[Dreizler 
\& Schuh(2001)]{2001ASPC..226...69D} Dreizler, S., \& Schuh, S.\ 2001, 12th European Workshop on White Dwarfs, 226, 69 


\bibitem[Dupree 
\& Raymond(1983)]{1983ApJ...275L..71D} Dupree, A.~K., \& Raymond, J.~C.\ 1983, \apjl, 275, L71 

\bibitem[Dupuis et al.(1993)]{1993ApJS...87..345D} Dupuis, J., Fontaine, 
G., \& Wesemael, F.\ 1993, \apjs, 87, 345 

\bibitem[Farihi et al.(2009)]{2009ApJ...694..805F} Farihi, J., Jura, M., 
\& Zuckerman, B.\ 2009, \apj, 694, 805 

\bibitem[Farihi et al.(2010)]{2010ApJ...714.1386F} Farihi, J., Jura, M., 
Lee, J.-E., \& Zuckerman, B.\ 2010a, \apj, 714, 1386 

\bibitem[Farihi et al.(2010)]{2010MNRAS.404.2123F} Farihi, J., Barstow, 
M.~A., Redfield, S., Dufour, P., \& Hambly, N.~C.\ 2010b, \mnras, 404, 2123 

\bibitem[Finley et al.(1997)]{1997ApJ...488..375F} Finley, D.~S., Koester, 
D., \& Basri, G.\ 1997, \apj, 488, 375 

\bibitem[Gnat et al.(2010)]{2010ApJ...718.1315G} Gnat, O., Sternberg, A., 
\& McKee, C.~F.\ 2010, \apj, 718, 1315 

\bibitem[Holberg et al.(1997)]{1997ApJ...484..871H} Holberg, J.~B., 
Barstow, M.~A., Lanz, T., \& Hubeny, I.\ 1997, \apj, 484, 871 

\bibitem[Holberg et al.(1998)]{1998ApJS..119..207} Holberg, J.~B., 
Barstow, M.~A., \& Sion, E.M. \ 1998, \apj, 119, 207 

\bibitem[Holberg et al.(1999)]{1999ApJ...517..841H} Holberg, J.~B., 
Bruhweiler, F.~C., Barstow, M.~A., \& Dobbie, P.~D.\ 1999, \apj, 517, 841 


\bibitem[Kilic et al.(2005)]{2005ApJ...632L.115K} Kilic, M., von Hippel, 
T., Leggett, S.~K., \& Winget, D.~E.\ 2005, \apjl, 632, L115 

\bibitem[Knauth et al.(2003)]{2003ApJ...592..964K} Knauth, D.~C., Howk, 
J.~C., Sembach, K.~R., Lauroesch, J.~T., 
\& Meyer, D.~M.\ 2003, \apj, 592, 964 

\bibitem[Koester 
\& Wilken(2006)]{2006A&A...453.1051K} Koester, D., \& Wilken, D.\ 2006, \aap, 453, 1051 

\bibitem[2009]{kout09} Koutroumpa, D., Lallement, R., Raymond, J. $\&$ Kharchenko, V., 2009, \apj, 696, 1517



\bibitem[Lallement(2009)]{2009SSRv..143..427L} Lallement, R.\ 2009, Space
Science Reviews, 143, 427

\bibitem[Landsman et al.(1993)]{1993PASP..105..841L} Landsman, W., Simon, 
T., \& Bergeron, P.\ 1993, \pasp, 105, 841 

\bibitem[Lehner et al.(2003)]{2003ApJ...595..858L} Lehner, N., Jenkins, 
E.~B., Gry, C., Moos, H.~W., Chayer, P., 
\& Lacour, S.\ 2003, \apj, 595, 858 

\bibitem[Marsh et al.(1997)]{1997MNRAS.286..369M} Marsh, M.~C., et al.\ 
1997, \mnras, 286, 369 

\bibitem[Melis et al.(2011)]{2011arXiv1102.0311M} Melis, C., Farihi, J., 
Dufour, P., Zuckerman, B., Burgasser, A.~J., Bergeron, P., Bochanski, J., 
\& Simcoe, R.\ 2011, arXiv:1102.0311 

\bibitem[Oegerle et al.(2005)]{2005ApJ...622..377O} Oegerle, W.~R., 
Jenkins, E.~B., Shelton, R.~L., Bowen, D.~V., 
\& Chayer, P.\ 2005, \apj, 622, 377 

\bibitem[Osterman et al.(2010)]{2010arXiv1012.5827O} Osterman, S., et al.\ 
2010, arXiv:1012.5827 

\bibitem[Peek et al.(2011)]{2011arXiv1104.5232P} Peek, J.~E.~G., Heiles, 
C., Peek, K.~M.~G., Meyer, D.~M., 
\& Lauroesch, J.~T.\ 2011, arXiv:1104.5232 

\bibitem[Redfield 
\& Linsky(2008)]{2008ApJ...673..283R} Redfield, S., \& Linsky, J.~L.\ 2008, \apj, 673, 283 (RLO8)

\bibitem[Saffer et al.(1998)]{1998ApJ...502..394S} Saffer, R.~A., Livio, 
M., \& Yungelson, L.~R.\ 1998, \apj, 502, 394 


\bibitem[Savage \& Lehner(2006)]{2006ApJS..162..134S} Savage,
    B.~D., \& Lehner, N.\ 2006, \apjs, 162, 134

\bibitem[Slavin 
\& Frisch(2002)]{2002ApJ...565..364S} Slavin, J.~D., \& Frisch, P.~C.\ 2002, \apj, 565, 364 

\bibitem[Slavin(1989)]{1989ApJ...346..718S} Slavin, J.~D.\ 1989, \apj, 346, 
718 


\bibitem[Vennes et al.(1996)]{1996ApJ...468..898V} Vennes, S., Chayer, P., 
Hurwitz, M., \& Bowyer, S.\ 1996, \apj, 468, 898 

\bibitem[Vennes et al.(1997)]{1997ApJ...480..714V} Vennes, S., Thejll, 
P.~A., Galvan, R.~G., \& Dupuis, J.\ 1997, \apj, 480, 714 

\bibitem[Vennes et al.(1998)]{1998ApJ...502..763V} Vennes, S., Christian, 
D.~J., \& Thorstensen, J.~R.\ 1998, \apj, 502, 763 

\bibitem[Vergely et 
al.(2010)]{2010A&A...518A..31V} Vergely, J.-L., Valette, B., Lallement, R., \& Raimond, S.\ 2010, \aap, 518, A31 

\bibitem[Welsh et 
al.(1999)]{1999A&A...352..308W} Welsh, B.~Y., Sfeir, D.~M., Sirk, M.~M., \& Lallement, R.\ 1999, \aap, 352, 308 

\bibitem[Welsh 
\& Lallement(2008)]{2008A&A...490..707W} Welsh, B.~Y., \& Lallement, R.\ 2008, \aap, 490, 707 

\bibitem[Welsh et al.(2010)]{2010ApJ...712L.199W} Welsh, B.~Y., Wheatley, 
J., Siegmund, O.~H.~W., \& Lallement, R.\ 2010, \apjl, 712, L199 

\bibitem[Welsh 
\& Lallement(2010)]{2010PASP..122.1320W} Welsh, B.~Y., \& Lallement, R.\ 2010, \pasp, 122, 1320 

\bibitem[Werner et al.(1998)]{1998ESASP.413..301W} Werner, K., Dreizler, 
S., Haas, S., 
\& Heber, U.\ 1998, Ultraviolet Astrophysics Beyond the IUE Final Archive, 413, 301 

\bibitem[Zuckerman et al.(2007)]{2007ApJ...671..872Z} Zuckerman, B., 
Koester, D., Melis, C., Hansen, B.~M., \& Jura, M.\ 2007, \apj, 671, 872 

\end{thebibliography}
\end{document}